\documentclass[useAMS,usenatbib,usegraphicx]{mn2e}

\title[Ultra-massive White Dwarfs]{On the Empirical Evidence for the Existence 
of Ultra-massive White Dwarfs}
\author[S. Vennes and A. Kawka]{S. Vennes$^{1}$\thanks{E-mail: svennes@fit.edu} and A. Kawka$^{2}$\thanks{E-mail: kawka@sunstel.asu.cas.cz}\\
$^{1}$Department of Physics and Space Sciences, 150 W. University Blvd, Florida Institute of Technology, Melbourne, FL 32901, USA\\
$^{1}$Astronomick\'y \'ustav AV \v{C}R, Fri\v{c}ova 298,CZ-251 65 Ond\v{r}ejov,
Czech Republic}

\begin{document}

\date{}

\pagerange{\pageref{firstpage}--\pageref{lastpage}} \pubyear{2008}

\title{On the empirical evidence for the existence of ultra-massive white dwarfs}

\maketitle

\label{firstpage}

\begin{abstract}
We re-examine the evidence for the existence of ultra-massive ($M>1.1M_\odot$) 
white dwarfs based on gravitational redshift of white dwarfs in common proper 
motion binaries or in clusters, on parallax measurements, on orbital solutions, 
and, finally, on the analysis of hydrogen line profiles. We conclude that the 
best evidence is largely based on the analysis of Balmer line profiles although 
the companion to the A8V star HR~8210 is a compelling case made initially using 
the large binary mass function and confirmed by an analysis of the Lyman line 
spectrum. The confirmation and identification of high-mass white dwarfs, more 
particularly non-DA white dwarfs, using parallax measurements may prove 
critical in establishing the population fraction of these objects and in 
constraining the high-end of empirical initial-mass to final-mass relations. 
The existence of a substantial population of ultra-massive white dwarfs 
supports the concept of a steeper initial-mass to final-mass relations linking 
$6\ M_\odot$ progenitors with $\ga 1.1 \ M_\odot$ white dwarfs.
\end{abstract}


\section{Introduction}

Ultra-massive white dwarfs, generally defined as having masses exceeding 
$1.1\ M_\odot$, remained relatively rare until deep spectroscopic surveys 
uncovered these intrinsically faint objects. The hot hydrogen-rich (DA) white 
dwarf GD~50 \citep{ber1991}, the carbon-rich white dwarf G35-26 
\citep{the1990}, and two DA white dwarfs (PG~1658+441 and PG~0136+251) from the 
Palomar-Green survey \citep{sch1992} were rare examples of this phenomenon. 
Interestingly, spectroscopic follow-up of extreme ultraviolet (EUV) surveys of 
the local ($d \la 100$ pc), hot white dwarf population managed to identify many 
new ultra-massive white dwarfs on the basis of a {\it large, spectroscopically 
determined surface gravity} ($\log{g}\ga 9$). Based on the {\it Extreme 
Ultraviolet Explorer} ({\it EUVE}) and ROSAT {\it Wide Field Camera} 
({\it WFC}) surveys, \citet{ven1996, ven1997b}, \citet{ven1999}, 
\citet{fin1997}, and \citet{ mar1997} added a dozen new objects to the 
population. More recently, a re-analysis of the Palomar-Green (PG) sample of 
DA white dwarfs \citep{lie2005} and a study of the white dwarf mass 
distribution in the Sloan Digital Sky Survey \citep[SDSS; ][]{kep2007} seem to 
generate similar yields of ultra-massive white dwarfs.

Are these objects products of single star evolution or double degenerate 
mergers? \citet{wei2000} reviews theoretical arguments in favour of an upper 
mass limit for white dwarf stars larger than the canonical upper limit of 
$1.1\ M_\odot$, and possibly as large as $\approx 1.3\ M_\odot$. A limit of 
$1.1\ M_\odot$ is generally assumed because carbon ignition in high mass cores 
would lead to thermonuclear runaway or core collapse. However, the effect of 
mass loss may alter the scenario and lead to the formation of a massive 
oxygen-neon-magnesium white dwarf \citep{nom1984}. \citet{gar1997} and 
\citet{rit1996} successfully evolved 9 and 10 $M_\odot$ stars, which, 
following an off-centre carbon ignition in partial electron degenerate 
conditions, generated oxygen-neon core white 
dwarfs with carbon-oxygen shells, and total masses of $1.15$ and 
$1.26\ M_\odot$, respectively. On the other hand, the merger scenario proposed 
for the origin of Type Ia supernovae \citep{yun1994} also generates 
ultra-massive white dwarfs. However, the merger process itself and the fate of 
these objects are uncertain \citep{seg1997}.

We re-examine the evidence for the existence of ultra-massive white dwarfs. 
Mass measurements are based on gravitational redshift measurements (\S 2), 
radius (parallax) measurements (\S 3), orbital parameters (\S 4), and surface 
gravity measurements (\S 5). The gravitational redshift measurements 
($\propto M/R$), radius measurements, and surface gravity measurements 
($\propto M/R^2$) are converted into mass measurements by adopting mass-radius 
relations for a variety of model structures. We adopted the models of 
\citet{ben1999} with a C/O core and zero metallicity, a helium mantle 
($\log{q}({\rm He}) = -2$), and with a {\it thick} ($\log{q}({\rm H})=-4$) 
hydrogen layer to describe hydrogen-rich DA white dwarfs or without a hydrogen 
layer to describe non-DA white dwarfs. In \S 5, we also present evidence that 
deep surveys of high-proper motion white dwarfs are now extending the 
ultra-massive population toward lower temperatures, and it is estimated that 
$\approx 10$\% of white dwarfs have masses in excess of  $1.1\ M_\odot$. We 
summarize and conclude in \S 6.

\section{Based on gravitational redshift: the CPMB sample and cluster members}

The massive white dwarf and member of the Pleiades LB~1497 \citep{weg1991b} 
remained a unique candidate for having a mass in excess of $1.0\ M_\odot$ and 
possibly in excess of $1.1\ M_\odot$ among stars with gravitational redshift 
measurements \citep{weg1991a, ber1995, rei1996}. However, applying the same 
techniques to white dwarfs in common proper motion binaries \citet{sil2001} 
proposed new ultra-massive white dwarf candidates (Table~\ref{tbl-1}). We 
demonstrate that this is not the case and that the ultra-massive white dwarf 
sequence presented by \citet{sil2001} and based on gravitational redshift 
measurements does not exist. We also re-evaluate the mass of LB~1497 and show 
that its case remains compelling.

\begin{table}
\centering
\begin{minipage}{80mm}
\caption{The CPMB Sample$^a$}
\label{tbl-1}
\begin{tabular}{ccc}
\hline
WD & CPMB & Types \\
\hline
0433$+$470 & LP~356-525/BD+26 730 & DA +dK5 \\
0738$-$172 & LP~783-3/LP783-2     & DZAQ+dM6 \\
0820$-$585 & LP~186-119/LP186-120 & DA+dM3 \\
1750$+$098 & G140-B1B/G140-B1A    & DC+dK2 \\
2323$-$241$^b$ & G275-B16A/G275-B16B &  F7+dK7 \\
\hline
\end{tabular} \\
$^a$ \citet{sil2001}. \\
$^b$ Not a white dwarf, see text.
\end{minipage}
\end{table}

\subsection{WD~0433$+$470: LP~356-525}

The hydrogen-rich (DA) white dwarf WD~0433$+$470 is part of a CPMB with the 
putative Hyades member BD+26~730. However, the presence of a cool, presumably 
old white dwarf in the Hyades poses a chronology problem that \citet{cat2008b} 
tentatively resolved by substituting a normal C/O core with a Fe core thereby 
shortening the cooling age of the white dwarf from 4 Gyr to 1Gyr. Their 
analysis is based on the Balmer line profile fits of \citet{ber2001} and 
\citet{zuc2003}. Adopting a straight average of these measurements, 
$T_{\rm eff}=5527\pm130$ K and $\log{g}=8.08\pm0.10$, we determined a mass 
$M=0.63\pm0.06\ M_\odot$ and absolute visual magnitude $M_V=14.78\pm0.30$ 
using the DA mass-radius relations of \citet{ben1999}. The resulting mass is 
well below the mass based on the gravitational redshift measurements of 
\citet{sil2001} $M=1.12\pm0.01\ M_\odot$. Moreover, the apparent ($V=15.81$) 
magnitude combined with the absolute visual magnitude for the low-mass solution 
corresponds to a distance modulus of $m-M=1.03\pm0.30$ ($d = 16\pm2$ pc) in 
agreement with the Hipparcos parallax measurement for BD+26~730, 
$\pi=56.0\pm1.2$ mas \citep{gou2004} or $d=17.9\pm0.4$ pc, and the white dwarf 
parallax measurement, $\pi = 60\pm3$ mas \citep{har1980} or $d=16.6\pm0.8$ pc. 
According to \citet{cat2008b}, the mass of an iron core white dwarf would be 
lower by $\approx 0.2\ M_\odot$ for the same temperature and surface gravity. 
The white dwarf radius inferred for this configuration would also decrease by 
18\% resulting in a closer distance for the white dwarf $d=13\pm2$ pc that is 
marginally consistent with the parallax measurements.
 
\subsection{WD~0738$-$172: LP~783-3}

The metal-rich hydrogen-poor white dwarf LP 783-3 (DZAQ6) is paired with the 
dM star LP 783-2. The parallax of the white dwarf, $\pi=112\pm3$ mas 
\citep{van1995}, translates into a distance modulus of $m-M=-0.25\pm0.06$ and, 
using $V=13.03$, we determined the absolute magnitude $M_V=13.28\pm0.07$.  

By fitting optical/infrared spectral energy distribution and by constraining 
the absolute magnitude to $M_V=13.31\pm0.05$ \citet{ber2001} determined  
$T_{\rm eff}=7710\pm220$ K and $\log{g}=8.09\pm0.03$. 

We revisited this result by computing a H/He model grid for effective 
temperatures $7000\le T_{\rm eff} \le 8500$ K and surface gravities 
$7.0\le \log{g} \le 8.5$, and a hydrogen abundance $\log{\rm H/He} = -3$ 
consistent with the measurements of \citet{ber2001}. The absolute magnitudes 
were calculated using the mass radius relations of \citet{ben1999} for non-DA 
stars. We adopted the temperature of \citet{ber2001}, $T_{\rm eff}=7710\pm220$, 
and the absolute magnitude $M_V=13.28\pm0.07$ and propagated the errors.
We determined a surface gravity $\log{g}=8.12\pm0.09$ that corresponds to a 
mass of $M=0.64\pm 0.06 M_\odot$. The mass of the white dwarf is {\it normal}. 

Adopting the distance modulus for the white dwarf ($m-M=-0.25\pm0.06$) we 
estimate an absolute luminosity for the red dwarf companion 
$M_V = 16.92\pm0.06$ that corresponds to a spectral dM6.5 \citep{kir1994}.
 
\subsection{WD~0820$-$585: LP~186-119}

\begin{figure}
\includegraphics[width=0.98\columnwidth]{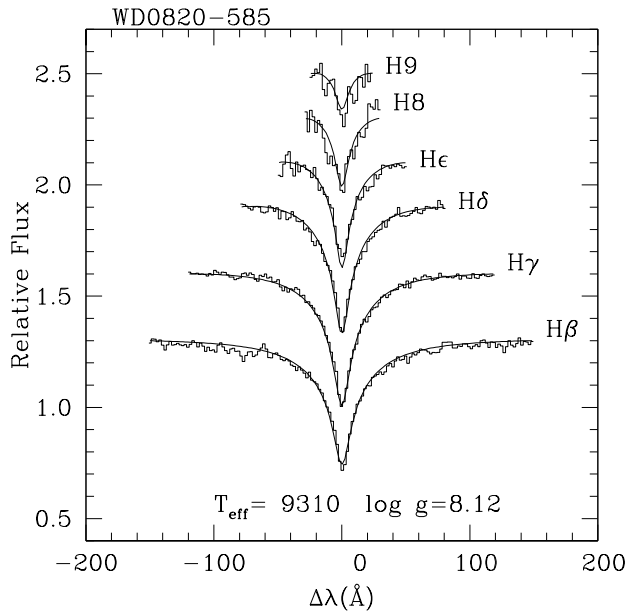}
\caption{Model atmosphere analysis of the DA white dwarf WD0820$-$585 (L186-119).}\label{fig1}
\end{figure}

The DA white dwarf (L186-119) is paired with a dM star (L186-120). We obtained 
a low-dispersion ($\approx$6 \AA) EFOSC2 spectrum from the ESO 
archives\footnote{Based on observations made with ESO telescopes at the 
La Silla Paranal Observatory under programme 078.D-0824.} and fitted the Balmer 
line profiles from H$\beta$ to H9 using our most recent pure hydrogen model 
atmosphere grid \citep{kaw2006}.
Figure~\ref{fig1} shows the results of our analysis. Using the DA mass-radius 
relations of \citet{ben1999}, the surface gravity and effective temperature 
measurements ($\log{g}=8.12\pm0.08$, $T_{\rm eff}=9310\pm70$ K) correspond to a 
mass $M=0.67\pm0.05\ M_\odot$ well below the mass of $M=1.10\pm0.03\ M_\odot$ 
estimated by \citet{sil2001} and based on their gravitational redshift 
measurement. The calculated absolute magnitude is $M_V=12.66\pm0.22$ and 
corresponds to a distance modulus $m-M=3.9\pm0.2$ ($d=60\pm6$ pc). 

\citet{sil2001} listed the spectral type of the companion as dM5. We revised 
this estimate using their V photometry as well as 2MASS JHK measurements 
\citep{cut2006}. The measured $V-J=4.0$ index corresponds to a $\approx$dM3.5 
($M_V=12.05$) spectral type \citep{kir1994} and the corresponding distance
modulus ($m-M=3.8$) confirms the physical association of the two stars and the 
{\it normal mass ($M\approx0.7\ M_\odot$) for the white dwarf component}.

\subsection{WD~1750$+$098: G140-B1B}

The DC white dwarf G140-B1B is paired with the K star G140-B1A (HD~162867). 
The optical/2MASS colour index $V-J=1.68\pm0.08$ of G140-B1A constrains the 
spectral type to K2 \citep{bes1988} and the absolute magnitude to 
$M_V=6.2\pm0.2$ \citep{dot2008}. Therefore, the distance modulus is 
$m-M=3.2\pm0.2$ and corresponds to a distance of $d=44$ pc. Adopting $V=15.72$ 
for the white dwarf \citep{egg1965}, the absolute luminosity is 
$M_V=12.5\pm0.2$. To determine the corresponding stellar parameters we computed 
a grid of pure He models at $9000\le T_{\rm eff} \le 11000$ K and surface 
gravities $8.0\le \log{g} \le 9.0$ and calculated the absolute magnitudes using 
the non-DA mass-radius relations of \citet{ben1999}. Combined with the colour 
index $V-I=0.178\pm0.060$ \citep{sil2001}, the absolute magnitude corresponds 
to $T_{\rm eff}=9200\pm600$ K and $\log{g}=8.14\pm0.25$ that translate into a 
mass of $M=0.66\pm0.16\ M_\odot$. The presence of H$\alpha$ in the white dwarf 
spectra is uncertain as \citet{sil2001} and \citet{weg1991a} claim H$\alpha$ 
detections while \citet{rei1996} does not report the detection. The absence of  
H$\alpha$ would invalidate the white dwarf gravitational redshift measurement 
of \citet{sil2001}. However, the coincident proper motions and the {\it normal 
mass for the white dwarf}, which implies similar distance moduli for the two 
stars, strongly favour a physical association for the pair.

\subsection{2323$-$241: G275-B16A}

\citet{sil2001} estimated a white dwarf gravitational redshift 
$v_g=132\pm1$ km~s$^{-1}$ that translated into a mass $M=1.19\pm0.01\ M_\odot$. 
This conclusion is based on the assumption that the star is a white dwarf part 
of a common proper motion binary and that the velocity differential is entirely 
caused by the gravitational redshift of the white dwarf. The companion is a 
late type star. We now re-examine the evidence. 

\begin{figure}
\includegraphics[width=0.98\columnwidth]{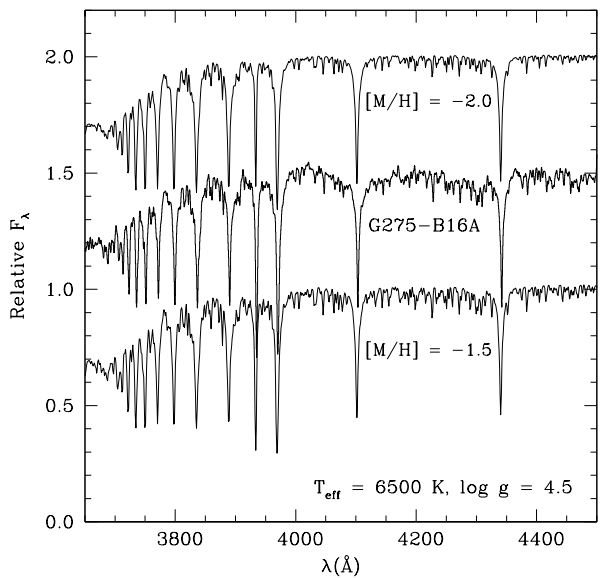}
\caption{ESO UVES spectrum of the F7 star G275-B16A (shifted up by 0.5 in 
relative flux units) compared to low-metallicity models, [M$/$H]$=-1.5$ and 
[M$/$H]$=-2.0$ (shifted up by 1.0 in relative flux units).}
\label{fig2}
\end{figure}

First, Figure~\ref{fig2} shows a comparison between two model spectra and the 
UVES spectrum from the ESO archives\footnote{Based on observations made with 
ESO telescopes at the La Silla Paranal Observatory under programme 165.H-0588.} 
of G275-B16A. The model spectra were obtained from the library of 
\citet{mun2005}\footnote{Available at http://archives.pd.astro.it/2500-10500/.},
which is based on Kurucz models. Models and data are presented at a resolution 
of 1~\AA. The Balmer line series clearly extends to H15 and implies a normal 
main-sequence surface gravity and a low metallicity. Indeed, we estimate 
[M$/$H]$\approx-1.5$ and $T_{\rm eff}=6500$ K ($\log{g}=4.5$). Therefore, the 
star G275-B16A is not  a white dwarf but a low metallicity main sequence star.  

Next, the distant modulus and proper motion measurements
suggest that the pair is coincidental. Based on 2MASS measurements and V 
magnitudes \citep{sil2005} we estimate $V-J=0.99$ and $2.43$ for G275-B16A and 
G275-B16B, respectively.  The colours correspond to spectral types of F7 and 
K7, respectively \citep{bes1988}. Using the low metallicity 
([M$/$H]$\approx-1.5$) models of \citet{dot2008} for the F star, the colour 
index $V-J=0.99$ corresponds to an absolute magnitude $M_V=5.2$ 
(or $T_{\rm eff}=6200$ K). A somewhat brighter absolute magnitude $M_V=4.9$ is 
obtained with the model at $T_{\rm eff}=6500$ K. Note that low-metallicity 
stars are fainter than their solar-metallicity counterparts. Using the 
solar-metallicity models for the K star we determined a temperature of 
$T_{\rm eff}=4200$ K and an absolute magnitude $M_V=8.0$. Therefore, the F 
star is at a considerably larger distance ($m-M=10.6$--$10.9$, 
$d=1.3$--$1.5$ kpc) than the K star ($m-M=7.7$, $d=350$ pc) and the pair is 
purely coincidental. 

This conclusion is supported by proper motion and radial velocity measurements 
that place the F star on a different course than the K star. \citet{zac2005} 
list the proper motions (in mas yr$^{-1}$) as $\mu_\alpha = 0\pm3$ and 
$\mu_\delta = -36\pm5$ for the F star, $\mu_\alpha = -5\pm5$ and 
$\mu_\delta = -64\pm5$ for the K star. Consequently, the difference in 
velocities, $v({\rm F\ star})-v({\rm K\ star})\approx 132$ km~s$^{-1}$
\citep{sil2001}, may no longer be interpreted as being due to the effect of 
gravitational redshift, but as being due to diverging space motions.

The F star space motion is characteristic of low-metallicity stars 
\citep{chi2000}. Adopting a distance of $1.4$ kpc the radial velocity 
($v_{\rm rad}=133$ km~s$^{-1}$) and proper-motion correspond, following 
\citet{joh1987}, to a Galactic velocity (in km~s$^{-1}$) vector 
$(U,V,W) = (143,-185,-138)$, which confirms its membership into the old disk 
or possibly the halo. 

\subsection{WD~0349+247: LB~1497}

LB~1497 is a member of the Pleiades and \citet{weg1991b} reported a 
gravitational redshift of $v_g=84\pm9$ km s$^{-1}$ for this DA white dwarf.  
\citet{ber1995} also resported a temperature $T_{\rm eff} = 31660\pm350$ K and 
a high gravity of $\log{g}=8.78\pm0.05$. Using the DA mass-radius relations of 
\citet{ben1999} and the temperature from \citet{ber1995} we obtain a mass of 
$M=1.03\pm0.04\ M_\odot$ based on the redshift measurement in agreement with 
the mass of $M=1.10\pm0.03\ M_\odot$ based on the surface gravity measurement. 
The weighted mass average is $M=1.075\pm0.024\ M_\odot$, close to the 
ultra-massive range.

\section{Based on parallax measurements}

Preliminary parallax information on G35-26 \citep{the1990}, GD~50 and 
PG~1658+441 was provided by \citet{dah1999}. Reading from Dahn's Figure 3, the 
absolute V magnitudes of these three objects are $12.9$, $11.5$, and $12.4$, 
respectively. Although, the original measurements are not listed the absolute 
magnitudes may be converted into masses using published effective temperatures 
and the mass-radius relations of \citet{ben1999}. Adopting 
$T_{\rm eff}=43200$ K for GD~50 \citep{ven1997b} and $T_{\rm eff}=30510$K for 
PG~1658+441 \citep{sch1992} we obtain masses of $M=1.24\ M_\odot$ and 
$M=1.29\ M_\odot$, respectively. These values are in good agreement with the 
spectroscopic masses \citep{sch1992, ven1997b}. 

\begin{figure*}
\includegraphics[width=0.98\textwidth]{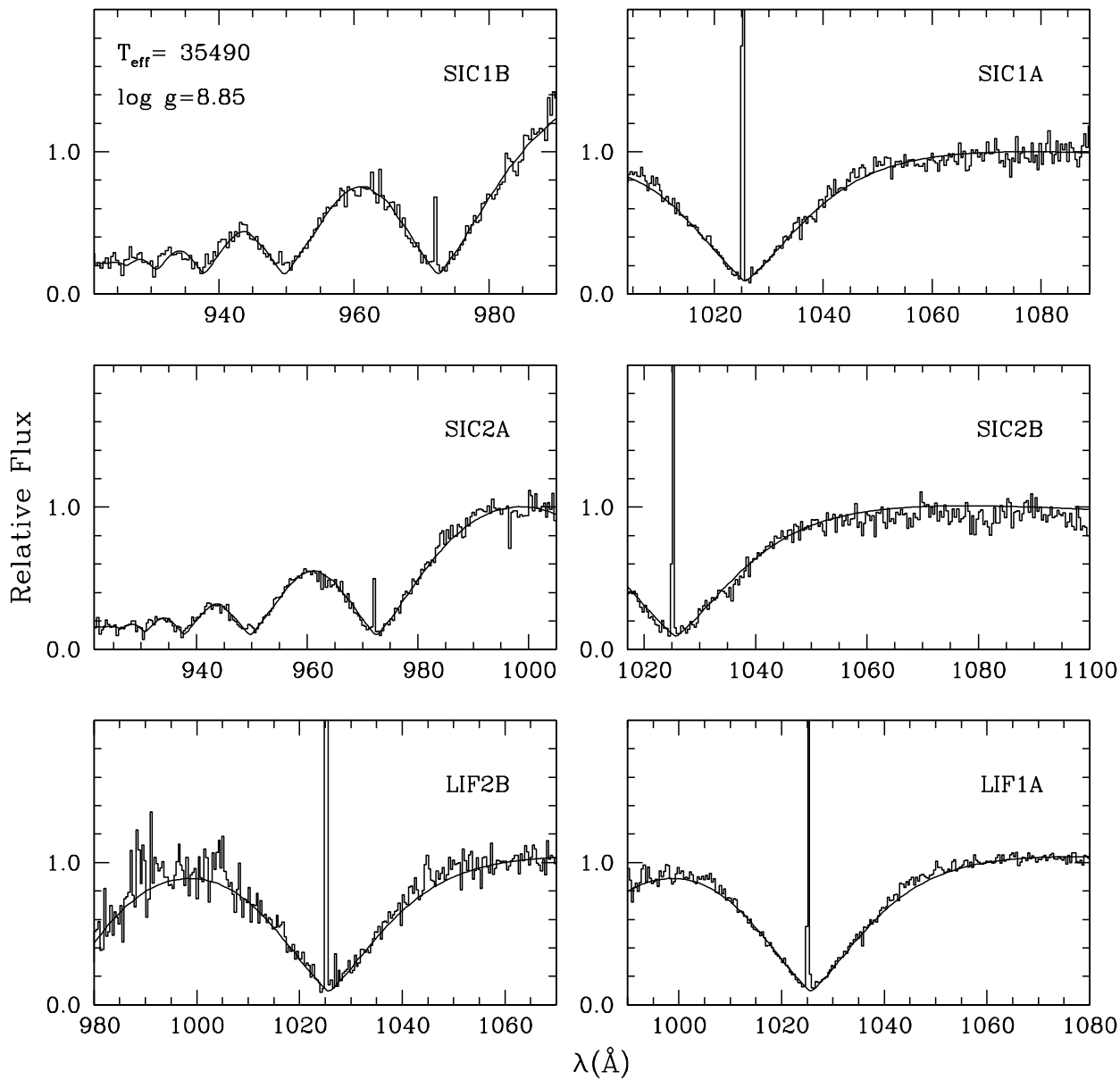}
\caption{Analysis of the Lyman line profiles of the DA white dwarf in HR~8210. 
Six separate {\it FUSE} spectroscopic channels are labelled and included in 
the analysis. Spectral ranges contaminated by geocoronal line emission were 
excluded from the analysis. The observed spectrum has been shifted by 
$-0.5$ \AA.}
\label{fig3}
\end{figure*}

Similarly, employing the effective temperature considered by \citet{the1990}, 
$T_{\rm eff}=14000$ K, and the non-DA mass-radius relations, the absolute 
visual magnitude of G~35-26, interpolated using a grid of helium model 
atmospheres at $7.0\le \log{g}\le9.5$ and $\log {\rm H/He}=-3$ and at a 
temperature of 14000 K, also suggest a high-mass of $M=1.16\ M_\odot$. This 
mass is somewhat lower than the spectroscopic mass obtained by \citet{the1990}, 
$M=1.2$--$1.33\ M_\odot$.

These comparisons and the analysis  of LHS~4033 \citep{dah2004}, and G47-18 and ESO439-26 \citep{ber2001} show that, although limited in scope, parallax measurements confirm the spectroscopic masses and the existence of a population of ultra-massive white dwarfs.

\section{Based on orbital solutions}

EUV surveys also uncovered a population of white dwarfs paired with early-type 
stars \citep{ven1998,bar2001}. Initial-mass to final-mass relations 
\citep{cat2008a} for white dwarf stars indicate that the progeny of early-type 
stars (A, B) would retain a mass above average and in excess of $1\ M_\odot$ 
for progenitors with $M\ga5\ M_\odot$. The massive subluminous companion of 
the A8V star HR~8210 is such a relevant case. Far and extreme ultraviolet 
spectroscopy uncovered a DA white dwarf companion to HR~8210 \citep{lan1993, 
bar1994,ven1998}. The mass function, $f(M_{WD})=0.219\pm0.004$ \citep{ven1998}, 
and the constraint placed on the orbital inclination ($i<88^\circ$) by the lack 
of an eclipse \citep{lan1995} restricts the mass of the white dwarf to 
$M({\rm WD})\ge1.19\ M_\odot$ if $M({\rm A8V})=1.6\ M_\odot$.

We confirmed the high mass for the white dwarf using {\it Far Ultraviolet 
Spectroscopic Explorer} ({\it FUSE}) spectra that cover the white dwarf 
spectral energy distribution from $\approx900$ \AA\ to $\approx1180$ \AA\ 
(Fig.~\ref{fig3}). The spectra (data id A0540909000) were obtained with 
{\it FUSE} on 2001 July 12 (UT) for a total exposure time of 4199 s using the 
LWRS aperture, and were processed using CalFUSE v3.2. The {\it FUSE} 
instrumentation is briefly described by \citet{moo2000}. 
We fitted the Lyman line series observed with {\it FUSE} with sets of grid of
models to constrain the effective temperature and surface gravity of the 
white dwarf. In total we used 4 sets of grids, which either include or
exclude the effect of Lyman satellite features and where line merging is 
treated either by using the formalism of \citet{ing1939} (hereafter IT) or of 
\citet{hum1988} (hereafter HH), which follows the treatment of \citet{hub1994}.
The latter is described in more detail in \citet{kaw2006} and \cite{kaw2007}.
Figure~\ref{fig3} shows the analysis of the Lyman line spectra using
the grid of model spectra that excludes Lyman satellites and used IT.
The best fit model to the spectra corresponds
to an effective temperature of $T_{\rm eff} = 35490\pm70$ and a surface
gravity of $\log{g} = 8.85\pm0.04$. We repeated the fit using a grid of models
that include Lyman satellites and use HH to obtain $T_{\rm eff} = 34350\pm100$
and $\log{g} = 9.04\pm0.05$. The grid of models that exclude Lyman satellites
and use HH result in a best fit with $T_{\rm eff} = 35000\pm80$ and 
$\log{g} = 8.75\pm0.03$. And finally the grid of models that include Lyman
satellites and use IT result in a best fit with $T_{\rm eff} = 34800\pm90$ and
$\log{g} = 9.13\pm0.04$. The series of analyses show that the two different
treatments of line merging produces a difference of $\approx 0.1$ in the
surface gravity, and that the inclusion of Lyman satellites increases the
measured surface gravity by $\approx 0.3$.
We find that the best agreement with \citet{ven1998} is obtained using models
that include the Lyman satellites however the fit is unsatisfactory as several
predicted features are stronger than observed. Similar difficulties were 
encountered by \citet{dup2003} in the analysis of Lyman line profiles of the
ultramassive DAp PG~1658$+$441.

Employing the mass-radius relations for DA white dwarfs \citep{ben1999},
we determined the mass and absolute magnitude of the white dwarf, 
$M=1.08 - 1.24\ M_\odot$ and $M_V=10.96 - 11.76$. \citet{ven1998} already
concluded that the distance modulus derived from the Hipparcos parallax of the
A8 star would be consistent with the predicted distance modulus for an
{\it ultra-massive white dwarf}. 

To our knowledge, the case of HR8210 is
rather unique with the possible exception of HD~209295. Based on their
analysis of the 3.1 day binary period, \citet{han2002} inferred the presence
of a companion to this A star. The measured UV excess and the constraints on 
the mass ($M > 1.04\ M_\odot$) suggest that the companion is a relatively hot
massive white dwarf, although the flux deficit in the TD-1 measurement at 1565 \AA\
conspires against this explanation.
The case of the 
putative white dwarf component of the triple system $\lambda$ Sco 
\citep{ber2000} was put to rest by \citet{uyt2004} who inferred the presence of 
a low-mass pre-main-sequence star rather than a hot ultra-massive white dwarf. 
No other ultra-massive white dwarfs were predicted based solely on orbital 
elements. It is worth noting that three other white dwarf companions to B type 
stars \citep[HR~2875, $\theta$ Hya, 16~Dra;][]{ven1997a, bur1999, bur2000} were 
uncovered from their extreme ultraviolet spectral signatures. Direct 
spectroscopic or orbital mass measurements are as yet unavailable.

\section{Based on Balmer line spectroscopy}

Figure 4 assembles the spectroscopic evidence gathered by \citet{ven1996}, 
\citet{ven1997b}, and \citet{ven1999} based on EUV surveys and as revised by 
\citet{ven2008} for a total of 158 objects, and \citet{kaw2006} and 
\citet{kaw2007} based on the New Luyten Two-Tenths (NLTT) proper-motion 
catalog (66 objects). In the latter, using proper-motion as a proxy for 
distance, the white dwarf selection probability is primarily a function of 
{\it distance} and the NLTT survey is primarily volume limited. Due to the 
particular morphology of the local interstellar medium and the so-called local 
bubble \citep[see][]{red2008}, \citet{ven1997b} argued that the white dwarf 
selection in EUV surveys is confined to a volume of radius $\la 100$ pc and 
characterized by low interstellar medium column density that does not hamper 
the EUV selection. They show that a white dwarf formation rate of 
$0.7-10\times10^{-12}$pc${-3}$ yr$^{-1}$ is sufficient to account for all hot 
white dwarf stars detected within 100 pc. Incidentally, \citet{lie2005} 
determined a formation rate of $0.6\times10^{-12}$pc${-3}$ yr$^{-1}$ for the 
DA white dwarfs in the PG survey. On this account, one may consider the EUV 
surveys to be volume limited as well and that few hot white dwarfs, 
irrespective of their radii (hence masses), were missing from within this 
volume. Interestingly, the yield in ultra-massive white dwarfs is very similar 
in both NLTT, 4 out of 66, and EUV, 17 out of 158, surveys. A total of 21 
objects out of 224 sampled have masses in excess of $1.1\ M_\odot$, and, 
therefore, close to 10\% of all white dwarfs maybe considered ultra-massive.

\begin{figure}
\includegraphics[width=0.98\columnwidth]{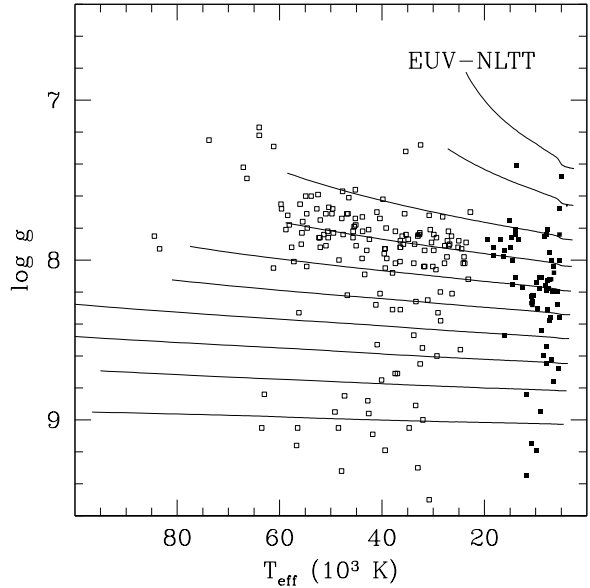}
\caption{Spectroscopically determined effective temperatures ($T_{\rm eff}$) 
and surface gravities ($\log{g}$) for the sample of EUV-selected DA white 
dwarfs (open squares) and a sample of hydrogen-rich DA white dwarfs from the 
NLTT survey (full squares; see \S 5). The measurements are compared to 
mass-radius relations for DA white dwarfs \citep{ben1999} for masses ranging 
from 0.3 (top) to 1.2 $M_\odot$ (bottom).}\label{fig4}
\end{figure}

\begin{figure}
\includegraphics[width=0.98\columnwidth]{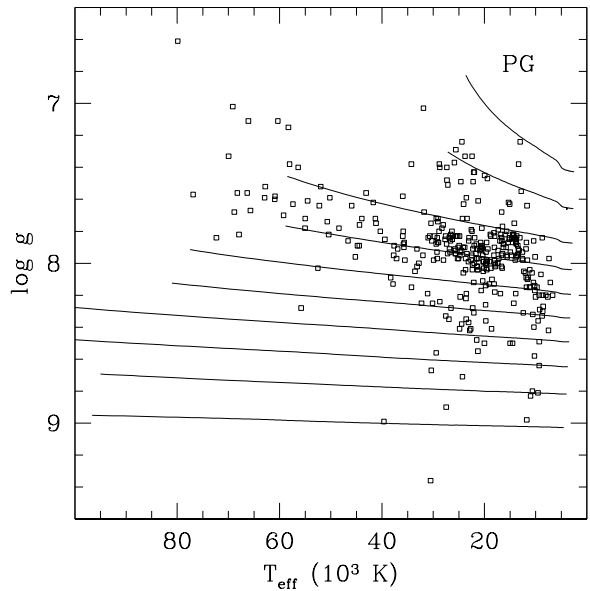}
\caption{Same as Figure~\ref{fig4}, bur for the Palomar-Green sample of \citet{lie2005}.}\label{fig5}
\end{figure}

\citet{lie2005} and \citet{kep2007} built white dwarf mass distributions based 
on the PG and SDSS surveys, respectively. Figure~\ref{fig5} shows the sample of 
347 DA white dwarfs analyzed by \citet{lie2005}. Only seven ultra-massive white 
dwarfs are extracted from this survey due to the relative faintness of these 
objects and the magnitude-limited survey strategy. Both ultra-massive white 
dwarfs hotter than 30\,000 K in the PG selection are also part of the EUV 
selection. However, by applying the $V/V_{\rm max}$ correction to the number 
counts they conclude, as previously established in the EUV-selected count, that 
$\approx 10$\% of white dwarfs maybe considered ultra-massive. A similar 
approach and conclusion was reached by \citet{kep2007}.

We now discuss the ultra-massive white dwarfs from the NLTT survey. 

\subsection{WD~0457$-$004: NLTT~14307}

\citet{kaw2006} measured a mass of $1.24\pm0.02\ M_\odot$. With a temperature 
of $10\,800$ K, the white dwarf lies outside the ZZ Ceti instability strip but 
it remains useful in helping define the location of its red edge at high masses.

\subsection{WD~1653$+$256: NLTT~43827}

\citet{kaw2006} measured a mass of $1.31\pm0.01\ M_\odot$. As in the case of 
NLTT~14307 and with a temperature of $11\,690$ K, the white dwarf is useful in 
helping define the location of the red edge of the instability strip at high 
masses.

\subsection{WD~1236$-$495: NLTT~31372}

The star is also know as LTT~4816 and is the most massive 
\citep[$M=1.11\pm0.02\ M_\odot$; ][]{kaw2007} pulsating DA star and the only 
known to date ultra-massive member of this class \citep{gia2005}.

\subsection{WD~1729$+$371: NLTT~44986}

The star is also known as GD~362 and is a peculiar ultra-massive 
($M=1.26\pm0.03\ M_\odot$) DAZ white dwarf \citep{gia2004, kaw2006}. The star 
is also perceived as harboring a disk of debris \citep{kil2005}.

However, \citet{zuc2007} recently obtained a high-dispersion, high
signal-to-noise-ratio spectrum of GD~362 that revealed a rich heavy-element
line spectrum, and one or possibly two \mbox{He\,{\sc i}} lines. The helium
abundance consistent with a weak \mbox{He\,{\sc i}}$\lambda$5876 \AA\ line has 
considerable effect on the effective temperature and surface gravity 
measurements. \citet{zuc2007} measured ($\log{\rm He/H}=1.1$) along with a
temperature $T_{\rm eff}=10540$ and $\log{g}=8.24$. The mixing of hydrogen
in an otherwise weakly detectable helium atmosphere has the effect of weakening
the upper Balmer lines and of mimicking the effect of a high gravity.
A parallax measurement is required to confirm the normal gravity
for this object.

\subsection{WD~2159$-$754: NLTT~52728}

The star is also known as  LTT~8816 and \citet{kaw2007} determined a mass of 
$M= 1.17\pm0.04\ M_\odot$. This estimate is supported by the peculiar radial 
velocity $v({\rm WD})=153\pm2$ km s$^{-1}$ \citep{max1999} that is much larger 
than normal. To illustrate this point we let the radial velocity of the star 
vary, while fixing its proper-motion to the observed value, 
$\mu=504$ mas yr$^{-1}$ and $\theta=277.9^\circ$ \citep{luy1976}, and compute 
the corresponding $(U, V, W)$ vector. By minimizing the difference between 
this vector and the local disk vector of \citet{chi2000} at 
$(U, V, W)_{\rm disk}=(0, -35, 0)$ we estimate the most probable radial 
velocity for the star, $v_{\rm rad} = 25$ km s$^{-1}$. Therefore, the excess 
$v({\rm WD})-v_{\rm rad}=128$ km s$^{-1}$ represents the most likely 
gravitational redshift for the white dwarf. This redshift translates into a 
mass of $1.17\ M_\odot$. Of course, without a velocity reference point 
(cluster or binary memberships) it is not possible to confidently determine 
the gravitational redshift but it is likely to be large. 

\section{Summary and conclusions}

We critically reviewed current evidence for the existence of ultra-massive 
white dwarfs. First, we demonstrated that the high-mass white dwarfs listed in 
\citet{sil2001} and presented as evidence for an ultra-massive white dwarf 
population in CPMB are in fact white dwarfs with normal masses 
($\approx 0.6$--$0.7\ M_\odot$), or, in the case of G275-B16A, a 
low-metallicity F star and possible halo member. The absence of any reliable 
candidates from this sample is puzzling considering the large number of objects 
identified using other methods. Improved radial velocity measurements of an 
enlarged sample of CPMB would help provide accurate gravitational redshifts and 
deliver the expected number of ultra-massive white dwarfs. 

The study of the peculiar DBAZ GD~362 exposes potential difficulties in
surface gravity, hence mass measurements based on hydrogen line profiles.
Although spectroscopically evanescent at an effective temperature
of 10\,000 K , helium was found to be the dominant
constituent in the atmosphere of that star. \citet{zuc2007}
demonstrated that the reduced hydrogen abundance had the consequence of
lowering the surface gravity measurement to almost a normal level.
Although helium lines are considerably stronger at temperatures in excess
of 15\,000 K, it is possible that some cooler ultramassive white dwarfs are
in fact helium dominated but with a normal mass.

Next, we examined the evidence based on parallax measurements and found that 
the small number of measurements, and the white dwarf masses inferred from 
these measurements are in good agreement with the spectroscopic masses. 
Table~\ref{tbl-2} lists spectroscopically-identified ultra-massive white dwarfs 
for which parallax measurements are desirable. 
Indeed, there remain some questions concerning the validity of our approaches, 
\citet{hum1988} or \citet{ing1939}, for Balmer and Lyman line merging at high
density and temperature. The effect of perturbers on upper energy levels is
essentially calibrated using normal gravity white dwarfs ($\log{g} = 8$) and
this calibration may not apply well at higher gravities potentially 
causing a systematic shift in mass measurements. However, a case-by-case
review of the few spectroscopically identified high-mass white dwarfs with 
parallax measurements shows good agreement between the two methods.
Table 2 lists the absolute visual magnitude 
and predicted parallax for each star based on parameters provided 
in the listed references. In addition to the calculated masses that we
determined using the mass-radius relations assuming a CO core, we have also
calculated mass estimates using mass-radius relations for white dwarfs with an 
ONe core \citep{alt2005}, which may be more appropriate for massive white 
dwarfs. The mass-radius relations for ONe cores, predict masses that are
systematically $\approx 0.02\ M_\odot$ lower than those predicted by CO 
mass-radius relations. For masses larger than $1.3\ M_\odot$, we used the
mass-radius relations of \citet{ham1961}.

We also confirmed the high mass for the hot white dwarf in the binary HR~8210. 
A high mass was initially implied by the binary mass function \citep{lan1993}. 
We fitted the Lyman line spectrum of the white dwarf and constrained the mass 
to be $M= 1.08 - 1.24\ M_\odot$ in agreement, in the upper mass range, with the binary parameters.

Finally, we show that the EUV-selected population of white dwarf stars is 
composed of $\approx 10$\% objects with masses in excess of $1.1\ M_\odot$. A 
similar yield was obtained by \citet{lie2005} based on the PG survey and by 
\citet{kep2007} based on SDSS {\it after large corrections} were applied due to 
the magnitude-limited nature of the samples collected. In particular, it should 
be noted that only seven objects out of 347 from the Palomar-Green sample, or a 
fraction of 2\%, met the criterion. By applying $V/V_{\rm max}$ corrections due 
to incompleteness at fainter magnitudes, the estimated fraction was 
re-evaluated at 10\% in agreement with the yield directly measured in the EUV 
selection. 

The origin of ultra-massive white dwarfs remains uncertain. Initial-mass to 
final-mass relations \citep{cat2008a} indicate that main sequence stars with 
masses in excess of $\approx 6\ M_\odot$ generate white dwarfs with masses in 
excess $1\ M_\odot$, a situation best illustrated by the massive white dwarf 
(LB~1497) member of the Pleaides. By re-evaluating available cluster data 
\citet{cat2008a} revised the final masses upward, and managed to reproduce the 
high-mass peak in both SDSS and PG empirical mass distributions. It is 
therefore possible that white dwarfs with masses in excess of $1.1\ M_\odot$ 
are the products of single star evolution and that the binary merger scenario 
may only apply to a minority of peculiar objects such as the fast rotating 
magnetic white dwarf WD~0325$-$857 \citep[see][and references therein]{ven2003}.

The existence of a substantial population of ultra-massive white dwarfs 
supports the concept of a steeper initial-mass to final-mass relations linking 
$6\ M_\odot$ progenitors with $\ga 1.1 \ M_\odot$ white dwarfs as proposed by 
\citet{cat2008a}. Ultra-massive white dwarfs in close binaries are also likely
Type Ia supernova progenitors \citep{par2007}.

\begin{table*}
\centering
\begin{minipage}{110mm}
\caption{NLTT and EUV spectroscopic sample of ultra-massive white 
dwarfs\label{tbl-2}}
\begin{tabular}{clccccc}
\hline
WD & Other names & Mass$^a$    & $M_V$ & Mass$^b$ & Predicted $\pi$ & Ref. \\
   &             & ($M_\odot$) & (mag) & ($M_\odot$) & (mas) & \\
\hline
0457$-$004 & NLTT~14307          & $1.24\pm0.02$ & 14.09 & $1.22\pm0.02$ & 58.8 & 1\\
1236$-$495 & NLTT~31372          & $1.11\pm0.02$ & 13.13 & $1.09\pm0.02$ & 66.7 & 2\\
1653$+$256 & NLTT~43827          & $1.31\pm0.01$ & 14.29 & $1.28\pm0.01$ & 30.3 & 1 \\
1729$+$371$^c$ & NLTT~44986          & $1.26\pm0.03$ & 14.54 & $1.24\pm0.03$ & 45.5 & 1\\
2159$-$754 & NLTT~52728          & $1.17^{+0.06}_{-0.07}$ & 14.30 & $1.14^{+0.05}_{-0.06}$ & 71.4 & 2 \\
0001$+$433 & EUVEJ0003$+$435     & $1.18\pm0.04$ & 11.19 & $1.16^{+0.04}_{-0.05}$ &  7.2 & 3\\
0136$+$251 & EUVEJ0138$+$253, PG & $1.13^{+0.04}_{-0.05}$ & 10.97 & $1.10\pm0.05$ &  9.9 & 3\\
0325$-$857 & EUVEJ0317$-$855     & $1.32\pm0.03$ & 12.64$^d$ & $1.32\pm0.03$ & 37.0$^d$ & 2, 4\\
0340$+$103 & 1RXSJ0343$+$1030    & $1.22^{+0.04}_{-0.05}$ & 11.01 & $1.20^{+0.05}_{-0.06}$ &  8.0 & 3\\
0346$-$011 & EUVEJ0348$-$009, GD~50 & $1.23\pm0.02$ & 11.49 & $1.21\pm0.02$ & 31.3 & 3\\
0440$-$038 & EUVEJ0443$-$037     & $1.14\pm0.07$ & 10.57 & $1.11\pm0.07$ &  5.4 & 3\\
0554$-$165 & 1RXSJ0557$-$1635    & $1.30^{+0.07}_{-0.12}$ & 11.86 & $1.28^{+0.09}_{-0.12}$ &  5.4 & 3\\
0630$+$200 & 1RXSJ0633$+$2001    & $1.22^{+0.04}_{-0.05}$ & 11.11 & $1.20^{+0.05}_{-0.06}$ &  6.9 & 3\\
0652$-$563 & EUVEJ0653$-$564     & $1.16^{+0.05}_{-0.06}$ & 11.40 & $1.13\pm0.06$ & 10.0 & 3 \\
0821$-$252 & 1RXSJ0823$-$2525    & $1.27\pm0.05$ & 11.83 & $1.25^{+0.04}_{-0.05}$ & 12.2 & 3\\
0914$-$195 & EUVEJ0916$-$197     & $1.27^{+0.06}_{-0.07}$ & 11.47 & $1.25^{+0.06}_{-0.08}$ &  6.8 & 3\\
1022$-$301 & 1RXSJ1024$-$3021    & $1.21\pm0.04$ & 11.63 & $1.19\pm0.05$ & 13.3 & 3\\
1529$-$772 & EUVEJ1535$-$774     & $1.26^{+0.03}_{-0.04}$ & 11.54 & $1.24^{+0.03}_{-0.04}$ & 10.6 & 3\\
1658$+$441 & EUVEJ1659$+$440, PG & $1.34^{+0.06}_{-0.14}$ & 12.83 & $1.33^{+0.07}_{-0.16}$ & 43.5 & 3\\
1711$+$668 & 1RXSJ1711$+$6645    & $1.22^{+0.03}_{-0.05}$ & 11.25 & $1.20^{+0.04}_{-0.05}$ &  6.8 & 3\\
1724$-$359 & EUVEJ1727$-$360     & $1.20^{+0.02}_{-0.04}$ & 11.66 & $1.17\pm0.03$ & 17.2 & 3\\
1740$-$706 & EUVEJ1746$-$706     & $1.17\pm0.04$ & 11.01 & $1.14\pm0.04$ &  7.9 & 3\\
\hline
\end{tabular}\\
$^a$ Using the mass-radius relations for a CO core \citep{ben1999}. For masses above $1.2 M_\odot$, mass-radius relations of
\citet{ham1961} were used. \\
$^b$ Using the mass-radius relations for a ONe core \citep{alt2005}. For masses above $1.3 M_\odot$, mass-radius relations of
\citet{ham1961} were used.\\
$^c$ Based on the model fit of Kawka \& Vennes (2006) assuming $He/H = 0$.
$^d$ Based on the distance modulus of the DA companion LB~9802 \citep{kaw2007}. \\
References: (1) \citet{kaw2006}; (2) \citet{kaw2007}; (3) \citet{ven2008}; (4) \citet{ven2003}.
\end{minipage}
\end{table*}

\section*{Acknowledgments}

S. V. is grateful for the hospitality and support of the Astronomical Institute 
at Ondrejov Observatory. A.K. acknowledges support from the Centre for 
Theoretical Astrophysics (LC06014). 
We are grateful to Jay Farihi for pointing out recent developments in the
study of GD~362.

\label{lastpage}

\end{document}